\documentclass[11pt,a4paper]{article}
\usepackage{amsmath}
\usepackage{amssymb}
\usepackage{amsthm}
\usepackage{palatino}
\usepackage{euler}
\usepackage{setspace}
\usepackage{xcolor}
\usepackage{hyperref}
\usepackage[german,british]{babel}
\usepackage[hang,flushmargin]{footmisc}
\usepackage[framemethod=TikZ]{mdframed}
\renewcommand{\vec}[1]{\boldsymbol{\mathbf{#1}}}
\newcommand{\reals}{\mathbb{R}}
\newcommand{\myatop}[2]{\genfrac{}{}{0pt}{}{#1}{#2}}

\begin{document}

\title{The Meaning and Facets of 
\emph{Symmetry}\\ in Physics}
\author{Domenico Giulini\\
\small Leibniz University of Hannover\\
\small Institute of Theoretical Physics\\
\small \texttt{giulini@itp.uni-hannover.de}}

\date{}

\maketitle

\noindent
{\footnotesize
This is an English translation of my contribution 
to the latest German edition \cite{Weyl:Symmetrie2017} 
of Hermann Weyl's classic text \emph{Symmetry}, 
which first appeared in English in 1952 
\cite{Weyl:Symmetry} and in 1955 in a 
German translation by the mathematician 
Lulu Bechtolsheim (1902--1989) 
\cite{Weyl:Symmetrie1955}. Our edition reproduces 
the 1955 German text, with substantially improved 
image quality, and includes annotations and 
commentaries by its editors, Erhard Scholz, 
Klaus Volkert and myself. We have also 
included the hitherto unpublished text 
``Similarity and Congruence: A Chapter in 
the Epistemology of Science'' (in English) 
from Hermann Weyl's literary estate at 
ETH Z\"urich. In this online version of 
my contribution to \emph{Symmetry}, I have 
added mathematical details in three shaded 
boxes labelled ``Technical Remark\,1'', 
``Technical Remark\,2'', and 
``Technical Remark\,3''.
\par}
\bigskip 

\begin{abstract}
\noindent
This contribution explores the various physical 
aspects of symmetry and places them in context 
from the perspective of theoretical physics. 
In particular, a distinction is drawn
between symmetries of individual configurations 
(solutions of equilibrium equations or equations 
of motion) and symmetries of the laws themselves. 
Likewise, a distinction is made between those 
operations that alter physical states 
and so-called gauge symmetries, which 
leave physical states unchanged and 
merely represent a redundancy in the mathematical description.
\end{abstract}

\section{Dynamical Symmetries in Physics}
Arguably, no mathematical concept has shaped modern physics more profoundly than \emph{symmetry}. Although 
``symmetry arguments'' have always 
played an important part in solving particular problems, it was only in the 
twentieth century that they were 
used systematically as a guiding heuristic principle in the discovery of 
new laws of nature. It is essential, however, to understand the different meanings of \emph{symmetry} in physics. 

For example, the elegant mathematical 
theory of crystal symmetries, discussed 
in detail by Weyl in this book 
\cite[p.\,120\,ff.]{Weyl:Symmetrie2017}, 
provides one physical application of the 
concept of symmetry. 
Ultimately, the structure of crystals is 
determined by the physical laws governing 
the interactions between atoms and 
molecules. The symmetry of crystals is 
therefore a case of the \emph{symmetry of 
solutions} of mathematical equations, for 
instance of the many-body Schr\"odinger 
equation of quantum mechanics, 
describing the constituents of the 
crystal. A solution is symmetric if operations corresponding to translations and/or rotations map it to itself; it is \emph{invariant}
under those operations. 
For this to be meaningful, however, the 
group in question must first be realised as a group of non-trivial 
operations on the set of solutions, for 
otherwise the question of whether a 
given solution is invariant has no mathematical 
meaning. The space of 
solutions must therefore carry a representation in the broadest sense.\footnote{By a ``representation in the broadest sense'' of a group $G$ on a set 
$M$ we always mean, here and in what 
follows, a non-trivial group homomorphism 
$G\rightarrow\mathrm{Bij}(M)$ into the group of 
bijections of $M$. If $M=V$ is a vector 
space, the term \emph{representation} often, 
though not always, refers to a \emph{linear 
representation}, that is, a group 
homomorphism $G\rightarrow\mathrm{GL}(V)$. 
Non-zero translations, for example, are affine but not linear transformations.}      

Let us illustrate this with an example from 
elementary physics: the combination of 
\textsc{Newton}'s law of gravitation
with his second law of motion. 
To this end, consider a system of $N$ point 
masses $m_a$, $a=1,\cdots,N$, moving in 
physical space under the influence of their 
mutual gravitational attraction. We assume that no additional forces act on these $N$ particles from external systems. 
We describe physical space with respect to an \emph{inertial frame}, so that 
the position of each point can be uniquely 
represented by three real numbers. For the 
$a$th particle, these three coordinates are 
$(x_a^1,x_a^2,x_a^3)$, which we also abbreviate 
as $\vec x_a$. The set of such triples of 
real numbers is denoted, as usual, by 
$\reals^3$, whose vector-space structure we 
use, with the standard operations 
of vector addition and scalar multiplication. 
In addition, we employ the Euclidean notion 
of distance, assigning to any two vectors 
$\vec x$ and $\vec y$ their distance 
$\Vert\vec x-\vec y\Vert$. This is given by 
$\sqrt{(x^1-y^1)^2+(x^2-y^2)^2+(x^3-y^3)^2}$ 
and is identified with the physical distance 
in space.\footnote{This identification of a 
mathematical notion of distance with an 
operationally defined physical one is not 
self-evident. It forms part of 
\textsc{Newton}ian mechanics, that is, part 
of the theory.} 

Under the influence of their mutual gravitational 
attraction, the $N$ particles move in a particular way. 
For the $a$th particle, this motion is described 
by a function $t\mapsto\vec x_a(t)$, called its 
\emph{trajectory}, which assigns to 
each instant $t$ of time the position of the 
$a$th particle in space. The motion of the 
entire system is determined once the trajectories 
of all particles are specified (with respect to the 
same absolute time $t$). These trajectories 
must satisfy the fundamental equation of 
\textsc{Newton}ian mechanics, 
$\text{force}=\text{mass}\times\text{acceleration}$. 
Since the force acting on the $a$th particle 
is the sum of the forces exerted by all the 
other particles, we must have 
\begin{subequations}
\label{eq:N-Teilchen}
\begin{equation}
\label{eq:N-Teilchen-a}
m_a\frac{d^2\vec x_a}{dt^2}=\sum_{\myatop{b=1}{b\ne a}}^N\vec F_{ab}\,.
\end{equation}
Here $\vec F_{ab}$ denotes the force exerted by the $b$th particle on the $a$th particle under consideration. By assumption, this is given by the gravitational force according to \textsc{Newton}'s law of gravitation: 
\begin{equation}
\label{eq:N-Teilchen-b}
\vec F_{ab}=-\gamma\,m_am_b\frac{\vec x_a-\vec x_b}{\Vert\vec x_a-\vec x_b\Vert^3}\,,
\end{equation}
\end{subequations}
where $\gamma$ denotes \textsc{Newton}'s gravitational constant.%
\footnote{We use the symbol $\gamma$ instead of the more usual $G$, since we wish to reserve the letter $G$ for a general group.}  
Substituting \eqref{eq:N-Teilchen-b} into \eqref{eq:N-Teilchen-a} gives $3N$ second-order differential equations for the $3N$ coordinates of the $N$ particles (three for each particle). These determine the motion completely once the positions and velocities of all particles are specified at a single, freely chosen instant.  
If we collect the $N$ vectors $\vec x_1,\cdots,\vec x_N$ in $\reals^3$ into a single vector $X:=(\vec x_1,\cdots,\vec x_N)$ in $\reals^{3N}$, we can combine the system of $N$ differential equations for the $N$ vectors $\vec x_a$ in $\reals^3$ arising from \eqref{eq:N-Teilchen} into a single differential equation for the time-dependent vector $X$ in $\reals^{3N}$:  
\begin{equation}
\label{eq:N-Teilchen-Kompakt}
E(X):=\frac{d^2X}{dt^2}-A(X)=0\,,
\end{equation}
where the ``acceleration function'' $A:\reals^{3N}\backslash\Delta\rightarrow\reals^{3N}$ is obtained by 
combining the $N$ $\reals^3$-valued functions 
$\sum_{b\ne a}\vec F_{ab}/m_a$. Here 
$\Delta\subset\reals^{3N}$ is the set of configurations 
for which $\vec x_a=\vec x_b$  for at least one pair of 
distinct particles $a\ne b$. We exclude these 
``collision configurations'' from the domain of $A$ for 
obvious reasons. 

When do we speak of \emph{symmetry} in connection with the physical situation just described? Perhaps the most important use of the term attributes symmetry to the dynamical law itself. What does this mean? Suppose we have a group $G$ that is represented on the set of trajectories $t\mapsto X(t)$. We simply write $(g,X)\mapsto g\cdot X$ for the action of the group element $g$ on the trajectory represented by the function $X:t\mapsto X(t)$. We call $G$ a symmetry group of the dynamical law \eqref{eq:N-Teilchen-Kompakt} if and only if every $g\in G$ maps solutions of \eqref{eq:N-Teilchen-Kompakt} to solutions of the \emph{same} equation (rather than a modified one). That is,
\begin{equation}
\label{eq:DefDynSymmetrie}
E(X)=0\,\Leftrightarrow\, E(g\cdot X)=0\,,\quad \forall g\in G\,.
\end{equation}
A distinct transformed solution cannot have the same 
initial conditions as the original, for otherwise 
uniqueness would clearly require the two to coincide. 
Initial data are not part of the dynamical law; 
specifying or determining them is the task of whoever 
applies it.

\section{Automorphisms of Space}
Examples of symmetries of the equations of motion 
\eqref{eq:N-Teilchen-Kompakt} are given by 
distance-preserving bijections of $\reals^3$. 
Any such transformation is either a translation, 
a linear orthogonal transformation, or 
a composition of the two.\footnote{Initially, we required only 
bijections, without assuming continuity or affine structure. 
Both follow from the preservation of distances. There are 
no ``wild'', discontinuous distance-preserving 
transformations; see \cite{Beckman.Quarles:1953}.}
Taken together, these form the Euclidean 
group $E_3$ of isometries of $\reals^3$, 
which is the semidirect product 
$\reals^3\rtimes O(3)$ of the group of 
all translations, here identified with the 
vector space $\reals^3$, and the group 
$O(3)$ of linear orthogonal 
transformations. If we endow space with 
an orientation, we restrict $O(3)$ to 
its subgroup $SO(3)$ of rotations, that 
is, orientation-preserving orthogonal transformations. $SO(3)$ is connected, and so is 
$E_3^+=\reals^3\rtimes SO(3)$, the group of 
\emph{Euclidean motions}.\footnote{%
Each element of $E^+_3$ can be thought of 
as the endpoint of a continuous one-parameter evolution that starts from the identity; 
hence ``motion''.} An element 
$g=(\vec a,R)\in\reals^3\rtimes SO(3)=E_3^+$ 
acts on space as   
\begin{equation}
\label{eq:DarstellungE3}
g\cdot\vec x=R\vec x+\vec a\,,
\end{equation}
where $R\vec x$ is the result of applying the rotation $R$ to $\vec x$. The requirement 
that \eqref{eq:DarstellungE3} be a 
representation (a homomorphism 
$E_3^+\rightarrow\mathrm{Bij}(\reals^3)$)   
reads 
\begin{equation}
\label{eq:DarstellungE3-2}
g_1\cdot(g_2\cdot\vec x)=(g_1g_2)\cdot\vec x\,.
\end{equation}
This requires 
\begin{equation}
\label{eq:E3-SemiDirectProd}
\bigl(\vec a_1,R_1\bigr)
\bigl(\vec a_2,R_2\bigr)=
\bigl(\vec a_1+R_1\vec a_2\,,\,R_1R_2\bigr)\,.
\end{equation}
This defines a group structure on the 
Cartesian product $\reals^3\times SO(3)$, called a semidirect product and written $\reals^3\rtimes SO(3)$ as above. This notation 
is meant to indicate that the underlying 
group structure differs in a particular
way from that of the direct product. 
In a direct product, both factors form normal subgroups. In a semidirect 
product, this is still true for one factor
(here $\reals^3$, the group of spatial translations), whereas the other factor 
(here $SO(3)$, the group of rotations) is still a subgroup but is not normal. 
This means that there is no single, 
distinguished subgroup of spatial rotations 
within $E_3^+$, but a whole class of subgroups, 
any two of which are related by conjugation with a translation. The subgroups in this class have different centres of rotation: points in space fixed by every element of the corresponding subgroup.

The affine 
representation of $E^+_3$ on $\reals^3$ extends 
in a natural way to an affine representation 
on $\reals^{3N}$, simply by applying each 
group element from $E^+_3$ simultaneously to 
each of the $N$ copies of $\reals^3$ in $\reals^{3N}$, as described in 
\eqref{eq:DarstellungE3}. It is now easy to 
see that the action of the group $E^+_3$ on 
the configuration space $\reals^{3N}$ of $N$ 
particles, implemented in this way, 
constitutes a symmetry in the above sense 
of the \textsc{Newton}ian equations of 
motion for gravitation.

\section{Automorphisms of Spacetime}
We thus see that, with an appropriate 
choice of representation, the geometric 
automorphisms of space become symmetries 
of certain physical laws of motion. 
These by no means exhaust the symmetries of the given dynamical law. As a generalisation of spatial automorphisms, 
one may consider spacetime automorphisms, which 
can likewise become symmetries of dynamical 
laws through the choice of suitable representations. 
In the present case, for example, we may translate points in space by an amount that depends linearly on time, while leaving time itself fixed: 
\begin{equation}
\label{eq:BoostTransformation}
(t,\vec x)\mapsto (t,\vec x+\vec vt)\,.
\end{equation}
Here $\vec v$ is a constant velocity 
vector whose three components specify 
the direction and magnitude 
of the spatial displacement in a unit interval of time. Transformations of the form \eqref{eq:BoostTransformation}
are called \emph{boosts}. 
In addition, we may perform pure translations 
in time while leaving all spatial coordinates 
fixed, so that
\begin{equation}
\label{eq:Zeittranslation}
(t,\vec x)\mapsto (t+b,\vec x)\,,
\end{equation}
where $b\in\reals$. The transformations 
\eqref{eq:DarstellungE3}, 
\eqref{eq:BoostTransformation}, and \eqref{eq:Zeittranslation}
together form the so-called \emph{Galilean group}, denoted by 
$Gal$, which is described by a total of ten continuously 
variable parameters: one for time translations ($b$), 
three for spatial translations ($\vec a$), three for 
boost transformations ($\vec v$), and three for 
rotations ($R$). We write elements $g\in Gal$ as 4-tuples $g:=(b,\vec a,\vec v,R)$ and spacetime points $p$ as pairs $p:=(t,\vec x)\in\reals\times\reals^3\cong\reals^4$. The action of 
$Gal$ on spacetime is then given by
\begin{equation}
\label{eq:ActionGal}
g\cdot p=
\bigl(b,\vec a,\vec v,R\bigr)\cdot
\bigl(t,\vec x\bigr)=
\bigl(t+b\,,\,
R\vec x+\vec vt+\vec a
\bigr)\,.
\end{equation}
This expression reflects a convention for interpreting the left-hand side. To specify an element of $Gal$ uniquely by a 4-tuple $(b,\vec a,\vec v,R)$, we must choose the order in which the corresponding transformations are composed and applied to a spacetime point. The relative order matters for 
those factors whose transformations 
do not commute. In \eqref{eq:ActionGal}
we first apply a rotation $R$, then a boost $\vec v$, then a spatial translation $\vec a$, and finally a time translation $b$. The group 
structure then follows from the representation 
condition
\begin{equation}
\label{eq:RepresentationProperty}
g_1\cdot(g_2\cdot p)=(g_1g_2)\cdot p\,,
\end{equation}   
which, together with \eqref{eq:ActionGal}, gives
\begin{equation}
\label{eq:MultGal-1}
\begin{split}
g_1g_2
&=\bigl(b_1,\vec a_1,\vec v_1,R_1\bigr)
 \bigl(b_2,\vec a_2,\vec v_2,R_2\bigr)\\
&=
\bigl(
b_1+b_2\,,\,
\vec a_1+R_1\vec a_2+b_2\vec v_1\,,\,
\vec v_1+R_1\vec v_2\,,\,
R_1R_2
\bigr)\,.
\end{split}
\end{equation}
Readers familiar with group theory will recognise that $Gal$ has the structure of an iterated semidirect product, as explained below:
\begin{equation}
\label{eq:GalSemiDir-1}
Gal\cong 
\bigl(\reals_b\times\reals^3_{\vec a}\bigr)
\rtimes
\bigl(\reals^3_{\vec v}\rtimes SO(3)\bigr)\,.
\end{equation}   
\noindent
\begin{mdframed}[linecolor=black, linewidth=1pt, roundcorner=3pt, backgroundcolor=gray!10]
{\fontsize{8pt}{10pt}\selectfont
\setlength\abovedisplayskip{4pt}
\setlength\belowdisplayskip{4pt}
\noindent
\textbf{Technical Remark 1.}
A brief digression into the structure of the Galilean group will put the preceding statements in context. Recall that a semidirect product $N\rtimes H$ 
of two groups $N$ and $H$ is characterised
by a homomorphism $\alpha: H\rightarrow Aut(N)$, 
$h\mapsto\alpha_h$, which determines the group structure on the set $N\times H$ as follows: 
\begin{equation}
\label{eq:DefSemiDirectProd}
(n_1,h_1)
(n_2,h_2):=\bigl(n_1\alpha_{h_1}(n_2)\,,
\,h_1h_2\bigr)\,.
\end{equation}
The identity element is $(e_N,e_H)$, where $e_N$ and $e_H$ are the identity elements of $N$ and $H$. Hence the inverse of $(n,h)$ is 
\begin{equation}
\label{eqSemiDirectProdInv}
(n,h)^{-1}
=\bigl(\alpha_{h^{-1}}(n^{-1})\,,
\,h^{-1}\bigr)\,.
\end{equation}
The group axioms can be verified using 
$\alpha_{h_1}\circ\alpha_{h_2}=\alpha_{h_1h_2}$ 
and $\alpha_h(n_1)\alpha_h(n_2)=\alpha_h(n_1n_2)$. 
The semidirect product reduces to a direct product if and only if $\alpha$ is trivial (the constant map 
onto the identity automorphism of $N$). For $Gal$, $N$ is the 
abelian group 
$\reals_b\times\reals^3_{\vec a}
\cong\reals^4$ of spacetime translations, 
whose group of (continuous!) automorphisms is $GL(\reals^4)$. 
[As an abelian group, $\mathbb{R}^n$ has a plethora 
of discontinuous automorphisms; as a vector space, 
however, it has only continuous automorphisms.]
The group $H$ is itself a semidirect 
product $\reals^3_{\vec v}\rtimes SO(3)$,
and the homomorphism $\alpha$ is 
\begin{equation}
\label{eq:SDPGal-1}
(\vec v,R)\mapsto
\alpha_{(\vec v,R)}
:=\begin{pmatrix}1&0\\
\vec v& R\end{pmatrix}\in GL(\reals^4)\,.
\end{equation}
One can now check that \eqref{eq:DefSemiDirectProd} 
with $n_{1{,}2}=(b_{1{,}2},\vec a_{1{,}2})$
and $h_{1{,}2}=(\vec v_{1{,}2},R_{1{,}2})$
yields \eqref{eq:MultGal-1}, using \eqref{eq:SDPGal-1}.
However, this is not the only way to 
give $Gal$ the structure of a semidirect product. Suppose that instead of 
$(b,\vec a,\vec v,R)$ we write 
$(\vec a,\vec v,b,R)$, and change the order of application accordingly: first the rotation $R$, then the time translation $b$, then the boost $\vec v$, and finally the spatial translation $\vec a$. Equation \eqref{eq:ActionGal} is then replaced by 
\begin{equation}
\label{eq:GalAltSDP-1}
\bigl(\vec a,\vec v,b,R\bigr)\cdot
\bigl(t,\vec x\bigr)=
\bigl(t+b\,,\,
R\vec x+\vec v(t+b)+\vec a
\bigr)
\end{equation}
and instead of \eqref{eq:MultGal-1} we get
\begin{equation}
\label{eq:MultGal-2}
\begin{split}
g_1g_2
&=\bigl(\vec a_1,\vec v_1,b_1,R_1\bigr)
 \bigl(\vec a_2,\vec v_2,b_2,R_2\bigr)\\
&=
\bigl(
\vec a_1+R_1\vec a_2-b_1R_1\vec v_2\,,\,
\vec v_1+R_1\vec v_2\,,\,
b_1+b_2\,,\,
R_1R_2
\bigr)\,.
\end{split}
\end{equation}
The latter shows that 
\begin{equation}
\label{eq:GalSemiDir-2}
Gal\cong 
\bigl(\reals^3_{\vec a}\times\reals^3_{\vec v}\bigr)
\rtimes
\bigl(\reals_b\times SO(3)\bigr)\,,
\end{equation} 
Thus $Gal\cong N\rtimes H$ is again a semidirect product. Here $N$ is the abelian normal subgroup of spatial translations and boosts, and $H$ is the direct product of time translations $\reals_b$ and rotations $SO(3)$. The homomorphism 
$\alpha:\reals_b\times SO(3)\rightarrow Aut(\reals^3_{\vec a}\times\reals^3_{\vec v})=GL(\reals^6)$,
which describes the action of $H$ on $N$ and replaces \eqref{eq:SDPGal-1},
is now given by 
\begin{equation}
\label{eq:SDPGal-2}
(b,R)\mapsto
\alpha_{(b,R)}
:=\begin{pmatrix}R&-bR\\
\vec 0& R\end{pmatrix}\in GL(\reals^6)\,.
\end{equation}
Again, one can check that \eqref{eq:DefSemiDirectProd} with
$n_{1{,}2}=(\vec a_{1{,}2},\vec v_{1,2})$
and $h_{1{,}2}=(b_{1{,}2},R_{1{,}2})$ 
gives \eqref{eq:MultGal-2} using \eqref{eq:SDPGal-2}.
\par}
\end{mdframed}

So far we have omitted spatial and temporal inversions. Including them introduces two further, mutually commuting involutions, $P$ and $T$. We may choose $P$ to be inversion about any point, for example the origin, 
so that $P(t,\vec x)=(t,-\vec x)$, and similarly choose $T$ to reverse time, for example $T(t,\vec x)=(-t,\vec x)$. Their actions on a trajectory $X$ are 
$(P\cdot X)(t):=-X(t)$ and $(T\cdot X)(t):=
X(-t)$. One can verify immediately, or see by inspection, that if $X$ satisfies the equation of motion \eqref{eq:N-Teilchen-Kompakt}, then so do  
$P\cdot X$ and $T\cdot X$.

\section{Similarities}
Similarities -- also known as homotheties -- 
are constant rescalings of the form 
\begin{equation}
\label{eq:Aehnlichkeiten}
\vec x\mapsto c\vec x\,,\quad c>0\,.
\end{equation}
These extend to $\reals^{3N}$ as $X\mapsto cX$. They are not symmetries of our equations of motion. Indeed, the first 
term in \eqref{eq:N-Teilchen-Kompakt} 
(involving a second derivative) scales 
proportionally to $c$, whereas the second term scales in proportion to $1/c^2$, as is 
immediately seen from \eqref{eq:N-Teilchen-b}. 

We can, however, make \emph{both} terms in \eqref{eq:N-Teilchen-Kompakt} scale homogeneously by rescaling time as well as space. The time rescaling must make the second derivative of $X$ scale in proportion to $1/c^2$ rather than $c$. This is achieved by allowing the scalings to act non-trivially on the time axis according to the combined prescription
\begin{subequations}
\label{eq:RaumZeitSkalierung}
\begin{alignat}{2}
\label{eq:RaumZeitSkalierung-a}
&\vec x&&\mapsto c\,\vec x\,,\\
\label{eq:RaumZeitSkalierung-b}
&t&&\mapsto c^{3/2}\,t\,.
\end{alignat}
\end{subequations}
The first equation \eqref{eq:RaumZeitSkalierung-a} again extends to $\reals^{3N}$ simply as $X\mapsto cX$. The scaling transformations with $c\in\reals_+$ form a group $G$, which we identify with $(\reals_+,\cdot)$. Let the group element $g\in G$ correspond to the positive real number $c\in\reals_+$. This group then acts on the set of trajectories $t\mapsto X(t)$ as follows: 
\begin{equation}
\label{eq:RaumZeitSkalierungGraph}
X\mapsto g\cdot X\,,\quad
(g\cdot X)(t):=c X\bigl(c^{-3/2}t\bigr)\,,
\end{equation}
and one readily verifies that $E(g\cdot X)(t)=c^{-2}\,E(X)(c^{-3/2}t)$, which immediately implies that this constitutes a dynamical symmetry in the sense of the definition given above. 
\medskip

\noindent
\begin{mdframed}[linecolor=black, linewidth=1pt, roundcorner=5pt, backgroundcolor=gray!10]
{\fontsize{8pt}{10pt}\selectfont
\setlength\abovedisplayskip{4pt}
\setlength\belowdisplayskip{4pt}
\noindent
\textbf{Technical Remark 2.}
The transition from \eqref{eq:RaumZeitSkalierung} to 
\eqref{eq:RaumZeitSkalierungGraph} is a 
special case of a general construction that defines group representations on spaces of maps from representations on their domains and target spaces. The simplest case is as follows. Let $M$ 
and $N$ be sets. Each carries a (general) 
representation (action) of the group $G$, 
$\phi:G\rightarrow\mathrm{Bij}(M)$, 
$g\mapsto\phi_g$, and 
$\psi:G\rightarrow\mathrm{Bij}(N)$, $g\mapsto\psi_g$. 
The induced representation 
$T: G\rightarrow 
\mathrm{Bij}\bigl(\mathrm{Map}(M,N)\bigr)$,
$g\mapsto T_g$, on the space of maps 
from $M$ to $N$ is then given by the following prescription ($\circ$ 
denotes composition of maps):
\begin{equation}
\label{eq:DefGroupMap-1}
T_g(X):=\psi_g\circ X\circ\phi_{g^{-1}}\,.
\end{equation}
This corresponds to mapping 
$\mathrm{Graph}(X):=\{(m,n)\in M\times N:
n=X(m)\}\subset M\times N$ via the combined 
action $\phi_g\times\psi_g$ of $G$ on 
$M\times N$. The image under $g$ of 
$\mathrm{Graph}(X)$ is then 
$\mathrm{Graph}(T_gX)$.

This construction can now be further 
generalised to the case in which 
$\psi$ is a \emph{twisted action} 
of $G$ on $N$, which means that it also 
depends on the base point in $M$. More 
precisely, instead of the homomorphism 
$\psi:G\rightarrow\mathrm{Bij}(N)$,
we now have a map $\psi:G\times M\rightarrow\mathrm{Bij}(N)$, 
$(g,m)\mapsto\psi_{(g,m)}$, satisfying
\begin{equation}
\label{eqDefGenGMap-1}
\psi_{(e,m)}=\mathrm{id}_N\,,\qquad
\psi_{\bigl(g_1,\phi_{g_2}(m)\bigr)}\circ\psi_{(g_2,m)}=\psi_{(g_1g_2,m)}\,.
\end{equation} 
A representation of $G$ on 
$\mathrm{Map}(M,N)$ is then given by 
\begin{equation}
\label{eq:DefGenGMap-2}
(T_gX)(m):=\psi_{\bigl(g,\phi^{-1}_g(m)\bigr)}
\Bigl(
 X\bigl(\phi^{-1}_g(m)\bigr)
\Bigr)\,.
\end{equation}
It indeed satisfies $T_{g_1}\circ T_{g_2}=
T_{g_1g_2}$ due to  \eqref{eqDefGenGMap-1}. 
Whereas \eqref{eq:DefGroupMap-1} suffices
to characterise, e.g., the action of the Lorentz
group on fields in special relativity, 
the more general law \eqref{eq:DefGenGMap-2}
is needed in general relativity and, in fact, also in Galilean relativity.  This is 
because boost transformations $(\vec v)$
act on particle trajectories $\vec x(t)\in\reals^3$ by shifting the value 
$\vec x$ by an amount depending on the 
argument $t\in\reals$. In this case 
$M=\reals_t$, $N=\reals^3_{\vec x}$, 
$\phi_{(b,\vec a,\vec v,R)}(t)=t+b$,
$\psi_{(b,\vec a,\vec v,R,t)}(\vec x)=
R\vec x+\vec vt+\vec a$. Hence
\eqref{eq:DefGenGMap-2} reads 
$(T_{(b,\vec a,\vec v,R)}\vec x)(t)=
R\vec x(t-b)+\vec v(t-b)+\vec a$. 
This specifies precisely how the Galilean group is represented 
on the set $\mathrm{Map}(\reals_t,\reals^{3}_{\vec x})$
and, correspondingly, on the set  $\mathrm{Map}(\reals_t,\reals^{3N}_{X})$ of general 
$N$-particle trajectories. It will then 
also be represented on the subset of 
\emph{kinematically possible} trajectories,
which, in view of \eqref{eq:N-Teilchen-Kompakt}, may be taken to be the 
piecewise twice differentiable elements 
of $\mathrm{Map}(\reals_t,\reals^{3N}_{X})$.  
For a given dynamical law, such as \eqref{eq:N-Teilchen-Kompakt}, the action is a symmetry if it maps the proper subset of \emph{dynamically possible trajectories} within the larger set of \emph{kinematically possible trajectories} to itself.
\par} 
\end{mdframed}

\section{Symmetry versus Covariance}
\label{sec:SymmetrieKovarianz}
Instead of rescaling time, we might have considered 
multiplying all masses $m_a$ by $c^3$ 
in order to achieve linear homogeneity in 
$c$, that is, $E(cX,c^3m_a)=c\,E(X,m_a)$, 
so that from $E(X,m_a)=0$ one may at least 
infer $E(cX,c^3m_a)=0$. This is not, however, a symmetry of the equation: the equation with masses $c^3m_a$ differs from the one with the original masses $m_a$. The same applies 
to the possibility of multiplying the 
gravitational constant $\gamma$ by $c^3$ 
in order to achieve homogeneity, that is,
$E(cX,m_a,c^3\gamma)=c\,E(X,m_a,\gamma)$, 
so that from $E(X,m_a,\gamma)=0$ one may 
infer $E(cX,m_a,c^3\gamma)=0$. Here again, 
one is relating \emph{different} equations 
of motion, provided that $\gamma$, or the masses 
$m_a$ in the preceding example, remain 
non-dynamical. There are, however, generalised 
theories of gravitation in which $\gamma$ is 
promoted to a dynamical field, 
meaning that it can in principle vary across 
spacetime and obeys its own equation of motion.
We shall not pursue this possibility here. 

These considerations reveal an important general feature of dynamical equations. Such equations typically 
involve, in addition to the dynamical 
quantities---collectively denoted here by 
$X$---further quantities that the equation neither constrains nor is intended to determine. Rather, these quantities define the equation itself. 
In physics, these are referred to as 
\emph{background fields} or \emph{absolute 
structures}. In a broad sense, parameters 
such as the masses $m_a$ in the above 
example also belong to this class. We then 
write the equations of motion in the form 
$E(X,H)=0$, where $H$ collectively denotes 
the non-dynamical quantities. Now, it 
makes a crucial difference whether one 
speaks of a dynamical symmetry in the sense 
of \eqref{eq:DefDynSymmetrie}, that is, in more explicit notation,
\begin{equation}
\label{eq:ProperDynSymmetrie}
E(X,H)=0\Leftrightarrow E(g\cdot X,H)=0\,,\quad \forall g\in G\,,
\end{equation}
or of a merely formal \emph{covariance} 
or \emph{form invariance} of the family of 
equations parametrised by $H$, under 
simultaneous transformation of the equation 
itself:
\begin{equation}
\label{eq:DynKovarianz}
E(X,H)=0\Leftrightarrow E(g\cdot X,g\cdot H)=0\,,\quad \forall g\in G\,.
\end{equation}  
The latter requirement can almost always be met by explicitly including the structures 
parametrised by $H$ in the equations and choosing 
the action of the group $G$ on these structures 
appropriately. Examples of this were seen above 
in connection with scaling, where we included 
either the masses $m_a$ or the gravitational 
constant $\gamma$ among the arguments of $E(\cdot)$ 
and scaled them accordingly so as to achieve 
covariance. By contrast, \eqref{eq:ProperDynSymmetrie} 
expresses a substantive symmetry claim. The symmetry group is 
the subgroup of the covariance group whose 
elements leave the non-dynamical quantities 
$H$ invariant. Enlarging the covariance 
group by introducing additional non-dynamical 
structures does not change the actual symmetry 
group. The latter can only change if one 
alters the status of the non-dynamical 
quantities and promotes them to dynamical 
ones, which is accompanied by the addition 
of further dynamical equations to those 
already present. The reverse process, 
in which formerly dynamical quantities are 
declared non-dynamical, and which is 
accordingly accompanied by a reduction 
of the symmetry group, will be discussed 
in connection with spontaneous symmetry 
breaking. 

\section{Internal Symmetries}
The examples considered so far still do not exhaust the symmetries of equation 
\eqref{eq:N-Teilchen-Kompakt}. In some cases, there are further symmetries 
that are not induced by transformations of space 
or spacetime. These are referred to as 
\emph{internal symmetries}. 

As an example, consider the special case 
of $N$ particles of \emph{equal} mass, 
that is, $m_1=m_2=\cdots=m_N=:m$. The group 
$S_N$ of permutations of $N$ objects is then 
represented linearly on $\reals^{3N}$ simply 
by permuting the positions of the particles:
$X=(\vec x_1,\cdots,\vec x_N)\mapsto 
(\vec x_{\sigma(1)},\cdots,\vec x_{\sigma(N)})$
for each $\sigma\in S_N$. These transformations 
are also symmetries in the above sense, since 
the transformed solutions are, as one readily 
verifies, again solutions that must be regarded 
as distinct from the original ones, at least 
if we suppose that further attributes distinguish 
the particles despite their equal masses. 

If we did not make this assumption, that is, 
if we assumed the particles to be indistinguishable, 
then the transformed solution would have to be 
regarded as identical to the original one, and 
we would have, instead of a dynamical symmetry 
of the equations of motion, merely a symmetry 
of the solution. In general, a symmetric solution is a fixed point of the 
dynamical symmetries in the space of solutions, 
that is, a solution for which $g\cdot X(t)=X(t)$ 
holds for all $g\in G$. If $G$ is the dynamical 
symmetry group, then solutions will in general 
not be symmetric at all, so that 
$g\cdot X(t)=X(t)$ holds only for the identity 
element $g=e$, or at most for all $g$ in a 
proper subgroup $G'\subset G$.

\section{Symmetries and Distinguishability}
\label{sec:SymmUnterscheidbarkeit}
Although the two concepts of symmetry---the 
dynamical symmetry of an equation and the 
symmetry of a particular solution of that 
equation---are clearly distinct, the preceding discussion shows that distinguishing them raises subtle questions. In the present 
case, the issue is whether we regard the permuted particle configuration as distinguishable in principle from the original or as entirely identical to it. For the dynamics under consideration, 
this is immaterial, since it depends only on 
the masses, which we have assumed to be equal. 
Any other attribute that might serve to 
distinguish the particles does not enter 
actively into the dynamics discussed here, 
so that, on the basis of the given mathematical 
description, there is almost no reason to prefer either interpretation. We 
say ``almost'', because we have, after all, 
labelled the particle coordinates by the 
natural numbers between 1 and $N$, thereby 
distinguishing them. It therefore remains open whether we did so merely for mathematical convenience or because we knew of further physical attributes by which the particles could be distinguished. 

In the first case, mathematical convenience would have led us to begin with a redundant description of the system, 
assigning several mathematical representatives 
to one and the same physical state. A faithful, non-redundant description would have to use equivalence classes of states, each comprising the states related by the symmetry group. In physics, transformations that map 
different mathematical representatives of one 
and the same physical state (or observable) 
into one another are called \emph{gauge 
transformations} or \emph{gauge symmetries}. 
The group formed by them is called the 
\emph{group of gauge transformations}, or 
sometimes simply the \emph{gauge group}.\footnote{%
The term ``gauge group'' is sometimes misleading, since in physics it is also 
used to denote what mathematicians call 
the \emph{structure group}. For example, 
in particle physics one speaks of 
an ``$SU(3)$ gauge theory'' (for the strong 
interaction) and refers to the (finite-dimensional) structure group $SU(3)$ as 
the gauge group, whereas the group of gauge 
transformations is instead given by the 
(infinite-dimensional) group of all 
$SU(3)$-valued functions on spacetime 
(more precisely: $SU(3)$ bundle automorphisms 
covering the identity on spacetime).}

Gauge symmetries therefore differ fundamentally 
from other, proper symmetries in their physical, 
operational meaning. The latter always assert the 
invariance of some aspect, such as the outward geometrical form of an object, under an 
operation that \emph{unequivocally changes} 
the overall state of the object. The rotation 
of a cube by 90 degrees relative to the 
surrounding space and all its other contents 
reproduces only part of the state, namely that 
defined by the positions of its 
indistinguishable vertices, edges, and faces 
relative to the surrounding space and its 
contents. Any distinguishing feature of the 
vertices, edges, or faces, for instance 
differences in colour, immediately shows that 
the state of the cube has in fact changed, 
that is, that a non-trivial operation has been performed on it. If no distinguishing feature 
existed even in principle, one would 
not be justified in claiming that any 
operation other than the identity had been 
performed on the object, at least if one 
adopts \textsc{Leibniz}'s principle of the 
\emph{identity of indiscernibles}.\footnote{%
Compare, e.g., Gonzalo Rodriguez-Pereyra, 
``The Identity of Indiscernibles'', 
The Stanford Encyclopedia of Philosophy 
(Summer 2025 Edition), 
\url{https://plato.stanford.edu/archives/sum2025/entries/identity-indiscernible}.} 
However, asserting a symmetry under the 
identity operation is clearly tautological. 
In this sense, the unqualified assertion of a 
``perfect'' or ``exact'' symmetry is, 
strictly speaking, meaningless.  
What is usually meant instead is that the 
particular aspect under consideration does 
not itself break the symmetry. This does not 
mean that no further symmetry-breaking aspect 
exists: it is precisely such an aspect that 
renders the operation, and hence the symmetry 
claim, non-trivial.

\section{Gauge Symmetries}
In contrast to the ordinary concept of symmetry 
described above, a \emph{gauge symmetry} does 
not alter the physical state, which is why one 
often also speaks of \emph{gauge redundancy}. 
Here the attributed redundancy lies in the way mathematical structures or symbols describe physical reality. One may, for example, introduce fields 
such as the gauge potentials in electrodynamics, 
from which the directly measurable physical 
fields---here the 
electric and magnetic fields---can be obtained 
by differentiation. The resulting map 
from the set of gauge potentials to the set of 
electromagnetic fields is surjective but not 
injective. The preimage of a given electromagnetic field is precisely a \emph{gauge orbit}, that is, the set obtained 
by applying all elements of the gauge group to 
any chosen element of the preimage. The set of 
physically measurable fields can thus be 
identified with the set of equivalence classes 
of gauge potentials modulo gauge transformations. At the level of the equations of motion, gauge symmetry usually manifests itself in the failure of those equations to determine the motion completely. Some of the field components 
and/or particle coordinates are then 
unconstrained in their time evolution by the equations of motion; their time dependence may be prescribed freely without contradicting those equations.

One may now ask what deeper purpose a gauge-symmetric formulation of a physical theory serves if it amounts merely to redundancy in the mathematical model. This is not an easy question to answer.  
The answer has to do with the fact that certain 
further requirements are imposed on the 
dynamical laws of physics, described by terms 
such as \emph{locality}, \emph{coupling}, and 
\emph{relationality}, and that these requirements can often be met simultaneously only at the price of a gauge-symmetric formulation.

Here we shall only comment on relationality and its connection with gauge symmetry. Once again, take $N$ point masses in physical space and suppose that only their mutual distances are physically measurable and dynamically significant. Whereas we usually take the view that the configuration of such a system of $N$ points is uniquely characterised by the $3N$-tuple
$X=(\vec x_1,\cdots,\vec x_N)$ of real numbers, we now assign the status of physical degrees of freedom only to those quantities that are \emph{invariant} under translations and rotations. Rigid translations and rotations of the system as a whole are therefore not to correspond to changes of physical state. In this case, the configuration of the system is uniquely determined by specifying the $\frac{1}{2}N(N-1)$ relative distances $r_{ab}:=\Vert\vec x_a-\vec x_b\Vert$. For $N\leq 4$, these distances provide independent parameters, so that the number of relational degrees of freedom for systems of 2, 3, and 4 particles is given by 1, 3, and 6, respectively. For five or more particles, the laws of Euclidean geometry impose dependencies among the $r_{ab}$, reducing the number of independent degrees of freedom to $3(N-2)$. The number of relational degrees of freedom for $1,2,3,4$ and $N>4$ particles is thus given by $0,1,3,6$ and $(3N-6)$, respectively. The simple expression $3N-6$ therefore gives the correct number of degrees of freedom for $N\geq 3$ particles. It fails for $N=2$, since in that case there is exactly one degree of freedom, namely the relative distance.
\medskip\par
\noindent
\begin{mdframed}[linecolor=black, linewidth=1pt, roundcorner=5pt, backgroundcolor=gray!10]
{\fontsize{8pt}{10pt}\selectfont
\setlength\abovedisplayskip{4pt}
\setlength\belowdisplayskip{4pt}
\textbf{Technical Remark 3.}
More generally, a system of $N$ particles in $\reals^D$ has
$F_1=ND$ degrees of freedom. If the relative distances of the $N$ particles are fixed, so that the system of particles forms a rigid framework, then in the case of $N\geq D$ particles (which we assume here) there remain
$F_2:=\frac{1}{2}D(D+1)$ degrees of freedom. We may picture these as comprising the $D$ degrees of freedom for the position of the first particle, the $(D-1)$ degrees of freedom for the position of the second particle at fixed distance from the first, the $(D-2)$ degrees of freedom of the third particle at fixed distances from the first and second, and so on, 
down to the single remaining degree of freedom for 
the position of the $D$th particle; the subsequent particles contribute no further continuous degrees 
of freedom. The rigidity condition therefore implies
$F_3:=F_1-F_2$ independent constraints on the original $F_1$ degrees of freedom. The number $F_3$ of rigidity constraints differs from the number $F_4:=\frac{1}{2}N(N-1)$ of relative distances of $N$ particles by
$\Delta:=F_4-F_3=\frac{1}{2}(N-D)(N-D-1)$. This difference vanishes for $N=D$ or $N=D+1$, but not for
$N>(D+1)$, where it gives the number of dependencies among the relative distances mentioned in the text. In three dimensions, there are therefore
$\frac{1}{2}(N-3)(N-4)$ such dependencies for $N\geq 3$ particles.
\par\medskip}
\end{mdframed}
\medskip

The crucial point is this. Suppose a system of $N$ particles consists of two subsystems of $N'$ and $N''$ particles, respectively, with
$N',N''\geq 3$. Then, in the non-relational (absolute) description, the subsystems have $3N'$ and $3N''$ degrees of freedom, respectively, and the composite system has their sum, namely $3(N'+N'')$ degrees of freedom. In particular, the configuration of the composite system is completely determined by the configurations of the subsystems. This changes in the relational description. Here the subsystems have $(3N'-6)$ and $(3N''-6)$ degrees of freedom, respectively, whereas the composite system has $3(N'+N'')-6$ degrees of freedom, six more than the sum for the subsystems. The configuration of the composite system is no longer determined by the configurations of the subsystems, and this indeterminacy has six degrees of freedom, corresponding to the six parameters specifying the position and orientation of one system relative to the other. Accordingly, the composite system also has relational properties that do not reside within the subsystems but concern the subsystems \emph{in relation to one another}. In particular, a dynamical coupling of two subsystems cannot, in general, be expressed solely by using the relational configuration variables of each of the subsystems, 
but must also  involve their \emph{mutual} relational degrees of freedom.

The connection with gauge transformations is clear: in the relational description, 
the six-dimensional group $E_3$ of 
Euclidean isometries\footnote{Here we deliberately choose $E_3$ rather than $E^+_3$ so as not to retain an absolute sense of orientation.} is the group of 
gauge transformations represented on the 
set of absolute configurations. Two configurations lying on the same group 
orbit count as indistinguishable, 
so that the set of relational configurations 
is faithfully represented only by the 
quotient of the set of absolute 
configurations by the equivalence relation 
defined by the group action. In our case, 
the configuration spaces in the absolute 
description are
$\reals^{3N'}$, $\reals^{3N''}$, and $\reals^{3(N'+N'')}$, and one has
$\reals^{3(N'+N'')}=\reals^{3N'}\times\reals^{3N''}$. In the relational description, these become
$\reals^{3N'}/E_3$, $\reals^{3N''}/E_3$, and $\reals^{3(N'+N'')}/E_3$, and the last of these is \emph{not} equal or isomorphic to
$\reals^{3N'}/E_3\times \reals^{3N''}/E_3$ (even their dimensions differ). Thus, in the relational description, the position and orientation of a subsystem \emph{as such} are not meaningful specifications, whereas its position and orientation relative to another subsystem are perfectly meaningful.

To couple two subsystems dynamically in a relational theory, we generally need more than their respective internal relational quantities. The coupling may depend, for example, on the distance between the subsystems. If one nevertheless wishes to represent the coupling by a mathematical expression containing only variables that are (formally) assigned either to the one or to the other subsystem, then this must necessarily involve gauge-\emph{variant} (absolute) quantities whose combination is once again gauge-\emph{invariant}. Thus, for example, the gravitational interaction between a system of point masses $m'_1,\cdots, m'_{N'}$ with absolute positions
${\vec x}'_1,\cdots ,{\vec x}'_{N'}$ and another system of point masses $m''_1,\cdots, m''_{N''}$ with absolute positions
${\vec x}''_1,\cdots ,{\vec x}''_{N''}$ is characterised by their interaction energy
\begin{equation}
\label{eq:WW-Energie}
V=-\,\gamma\sum_{a=1}^{N'}\sum_{b=1}^{N''}\frac{m'_am''_b}{\Vert{\vec x}'_a-{\vec x}''_b\Vert}\,,
\end{equation}
which depends exclusively on relative distances and is therefore gauge-invariant, but can be expressed only with the aid of gauge-\emph{variant} variables of the subsystems.
      
Correspondingly, in modern gauge field theories, couplings to other systems are expressed through gauge-variant field variables. The coupling of the electromagnetic field to an electron, for example, is represented by the integral of the local product of the field's gauge-variant potential and the electron's gauge-invariant current density. Owing to the local conservation of electric current, this integral is itself gauge-invariant. The situation is similar in the gauge field theories of the other interactions (strong and weak), except that there, because of the non-\textsc{Abel}ian character of the gauge group, the current too is gauge-\emph{variant}.

\section{An Aside on the Philosophy of Space}
The contrast between local and relational features of physical systems offers an interesting perspective on the 
ontological status of physical space, a 
topic also taken up by \textsc{Weyl} in
\cite[p.\,29]{Weyl:Symmetrie2017}. There 
\textsc{Weyl} refers to \textsc{Kant}'s 
famous discussion of the existence of bodies in 
space that are isometric yet cannot be 
brought into coincidence by means of 
(continuous) motions, such as a right 
and a left hand.\footnote{This example is 
also discussed in Klaus Volkert's 
contribution. Note the terminology: 
\textsc{Kant} calls such pairs of bodies 
\emph{not congruent} or \emph{incongruent}, 
which does not, however, correspond to 
modern usage, according to which 
reflections---that is, isometries not 
continuously connected with the 
identity---are counted among the
congruences.}  

\textsc{Kant} repeatedly commented on a fact that evidently made a deep impression on him: two bodies that are intrinsically geometrically identical, that is, isometric, can nevertheless appear geometrically different in some sense that needs to be clarified. This happens when they cannot be brought into coincidence by a continuous one-parameter family of motions starting at the identity, as with two figures that are mirror images of one another. He first 
addressed this topic at considerable 
length in his pre-critical essay 
\emph{Von  dem ersten Grunde des
Unterschiedes der Gegenden im Raum} 
of 1768, where from the existence of incongruent but isometric bodies he 
infers the independent (that is, body-independent) existence of 
space~\cite{Kant:GegendenImRaum}. Only two years later, in 1770, he returned briefly to the topic, namely in his inaugural 
dissertation \emph{Von der Form der 
Sinnen- und Verstandeswelt und ihren 
Gr\"unden}, in which, now on the threshold of his critical period, 
he rather surprisingly states that 
``only by means of a kind of pure 
intuition can difference, namely
incongruence, be
noticed'' \cite{Kant:InauguralDiss}. 
The matter rested there until 
\textsc{Kant} took it up once more in 1783, 
two years after the publication of the 
\emph{Critique of Pure Reason}, now from within his new critical philosophy, namely in \S\,13 of the 
programmatic work intended for a wider 
readership, \emph{Prolegomena zu einer 
jeden k\"unftigen Metaphysik, die als Wissenschaft wird auftreten k\"onnen} 
(often simply called 
the ``Prolegomena''). Here it becomes especially clear how central this phenomenon, now called a ``paradox'', was to the development of his transcendental-philosophical conception of space and time 
\cite{Kant:Prolegomena} (\S\,13, 
pp.\,147--148):
\begin{quote}
\emph{\noindent
Those who still cannot free themselves 
from the notion that space and time are 
real determinations inhering in things in 
themselves may exercise their ingenuity on 
the following paradox, and, if they have 
tried in vain to resolve it, may at least 
for a few moments become free of prejudice 
and suspect that perhaps there may indeed be 
grounds for degrading space and time to mere 
forms of our sensible intuition.%
\footnote{German original: 
``Diejenigen, welche noch nicht 
von dem Begriffe loskommen k\"onnen,
als ob Raum und Zeit wirkliche Beschaffenheiten w\"aren, die den 
Dingen an sich selbst anhingen, k\"onnen 
ihre Scharfsinnigkeit an folgendem 
Paradoxon \"uben, und, wenn sie dessen 
Aufl\"osung vergebens versucht haben, 
wenigstens auf einige Augenblicke von 
Vorurteilen frei, vermuten, dass doch 
vielleicht die Abw\"urdigung des Raumes 
und der Zeit zu blo{\ss}en Formen unserer 
sinnlichen Anschauung Grund haben 
m\"ogen.''}
}
\end{quote}
In connection with the problem of space, 
\textsc{Weyl} cites only the last of the three works mentioned above. For the physicist, 
however, \textsc{Kant}'s first, pre-critical 
essay \cite{Kant:GegendenImRaum} offers a more tangible treatment. Let us explain why.
Since \textsc{Einstein}'s formulation of the special theory of relativity, and especially following its mathematical formulation by \textsc{Minkowski}~\cite{Minkowski:1909}, modern physics has indeed ascribed to 
spacetime (not to space alone) an ontological status akin to that of \textsc{Newton}'s absolute space. General relativity does not alter this, for although the geometry of spacetime is influenced by the local distribution of the matter that fills it, its \emph{existence} is not tied to the presence of matter. Physicists today thus speak quite naturally of ``vacuum solutions'' of the \textsc{Einstein} field equations, that is, spacetimes without any material content whatsoever, or of gravitational waves that propagate from one system (a source) to another (a detector), carrying energy, momentum and angular momentum and changing the metric structure of spacetime as they pass through otherwise entirely matter-free regions. All this presupposes a 
``real determination'' (``wirkliche 
Beschaffenheit'') of spacetime that is  
obviously incompatible with a ``degradation''
(``Abw\"urdigung'')  of spacetime to a 
``mere form of our sensible intuition''. For 
this reason, \textsc{Kant}'s pre-critical 
engagement with the problem of space
(generalised to spacetime) seems to regain 
significance, at least in the present 
physical context, even though it might 
appear outdated from a philosophical point of 
view.

In \cite{Kant:GegendenImRaum}, \textsc{Kant} argues that the cause of the distinguishability of isometric but not continuously superposable bodies lies in their relation to the surrounding (physical) space, thereby securing its ontological status, in keeping with \textsc{Newton} and against the purely relational conception of \textsc{Leibniz}.
The mathematician \textsc{Kurt Reidemeister}
(1893--1971) analysed the logic of \textsc{Kant}'s 
argument and identified and listed its implicit assumptions~\cite[Ch.\,IV]{Reidemeister:RaumUndZahl}. In essence, \textsc{Kant} maintains that the opposite orientations of two otherwise isometric bodies, such as hands, must be grounded in a \emph{spatial} difference between them. He assumes that a single hand, considered on its own, is already either a left or a right hand, irrespective of the presence of its counterpart; compare \cite[p.\,29]{Weyl:Symmetrie2017}. There must therefore be a spatial property of one hand that the other lacks: ``being left-handed'', for example. But if we build ordinary Euclidean geometry upon the relational notion of distance, the preceding discussion already shows that the properties of a whole consisting of two oppositely oriented systems of points are not simply the union of their individual properties. \textsc{Reidemeister} showed that this applies in particular to orientation, emphasising the point in the following statement
\cite[Ch.\,IV, p.\,67]{Reidemeister:RaumUndZahl}: 
\begin{quote}
\emph{\noindent
Let $G',G''$ be two geometrical objects, and let $E'$ be the totality of all true geometrical statements about $G'$, $E''$ those about $G''$, and $E$ those about the ordered pair of objects $G',G''$. There are such pairs $G',G''$ for which the propositions of $E$ are not identical with the logical consequences of the propositions of $E'$ and $E''$.} 
\end{quote}
The distance between two points is already a property of this kind. A non-zero distance allows us to speak of \emph{two} points, even if they are intrinsically indistinguishable. Neither their being two nor their distance is a property that can be localised in either point. The same holds for the properties of \emph{parallelism} and \emph{orthogonality} (of straight lines).

\textsc{Reidemeister} confines his discussion to Euclidean geometry. From a more modern perspective, his response to \textsc{Kant} could be qualified by observing that whether two isometric figures can be brought into coincidence by continuous motion depends on the topology of the space in which the motion takes place. In particular, the properties of \emph{dimension} and \emph{orientability} play a role.

Consider dimension first. In the two-dimensional plane of a sheet of paper, an orientation can be defined by the ``two-legged figure'' $\rightharpoonup$, consisting of two legs of unequal length issuing from a common point. This figure cannot be moved continuously within the plane into $\rightharpoondown$, its reflection in the horizontal axis through the longer leg. It can, however, be brought into that position in the surrounding three-dimensional space. For example, we can rotate it continuously about the axis through its longer, horizontal leg from zero to 180 degrees, with the shorter leg temporarily leaving the plane of the paper. This example readily generalises to any dimension: an ``$n$-legged figure'' consisting of legs of pairwise unequal length issuing from one point and, for ease of visualisation, taken also to be pairwise orthogonal, can be moved continuously into an $n$-legged figure of opposite orientation if the motion takes place in $(n+1)$-dimensional Euclidean space, for example as a 180-degree rotation in a plane spanned by the new dimension and one of the legs (the latter thereby changing its ``direction''). To avoid this ambiguity, the dimension of the space must be fixed whenever one speaks of a ``motion in space''.
 
More interestingly, even at fixed dimension, a space may or may not be orientable. If it is not orientable---a possibility that is in no way especially exotic in modern differential geometry and topology---then it is possible, by continuous motion \emph{within} space itself, to change the orientation of a body or a figure. For an easily visualised two-dimensional example, imagine moving the two-legged figure $\rightharpoonup$ on our sheet of paper. Now modify the ``paper'' by joining its top and bottom edges, identifying the top left corner with the bottom right and the top right with the bottom left. This produces a M\"obius strip. One then sees immediately that a continuous translation once across the seam and back to the point of departure transforms the symbol $\rightharpoonup$ into $\leftharpoonup$. A further 180-degree rotation in the plane of the paper then yields $\rightharpoondown$.

One might therefore think that in non-orientable 
spaces it is altogether meaningless to ascribe 
an orientation to spatial figures, and that 
even the statement that two figures differ in 
orientation is not meaningful. This would seem 
to undermine \textsc{Kant}'s argument for the 
existence of space. Against this, however, one may 
object that orientation-reversing round trips 
cannot be realised along arbitrarily short paths.
Locally, space is always orientable, and it remains 
meaningful to ascribe an orientation to geometrical 
figures so long as they do not leave a certain local 
neighbourhood.%
\footnote{Since orientation-reversing closed curves 
cannot be continuously contracted to a point, the 
assignment of an orientation to a body would be 
meaningful at least so long as the body does not leave a 
simply connected region.} If bodies with locally 
different orientations could nevertheless be brought 
into coincidence by moving them over sufficiently 
long paths, we would have to infer the 
non-orientability \emph{of space}, which would 
thereby have reasserted its ontological status.

Thus \textsc{Kant}'s early, pre-critical engagement 
with the problem of space, suitably transferred to 
spacetime, has lost less of its relevance than one 
might initially suppose. \textsc{Weyl}'s position on 
\textsc{Kant}'s problem in 
\cite[pp.\,129--130]{Weyl:Symmetrie2017} is, in the 
first instance, a mathematical reformulation rather 
than a resolution. It remains to explain why the groups 
of geometrical and physical 
automorphisms differ and how that difference can be established. According to \textsc{Weyl}, the very statement of such a difference is possible only because, after abstraction from physical objects (matter), the group of dynamical symmetries remains 
as the group of physical automorphisms \emph{of space} 
\cite[p.\,129]{Weyl:Symmetrie2017}, namely as a subgroup of the bijections \emph{of space}, and can therefore be compared with the geometrical automorphisms, which likewise form a subgroup of the bijections. If, following \textsc{Weyl}, we \emph{define} the geometrical automorphism group as the normaliser of the physical automorphism group---specifying the ambient group in which the normaliser is taken---then, as \textsc{Weyl} observes, we need no longer be surprised that the physical automorphisms generally form a proper subgroup of the geometrical ones. That is true---at the level of group theory! But this group-theoretic reduction is possible only because space has been assumed to exist independently of matter as the medium on which \emph{both} groups are represented. From a 
\textsc{Kant}ian point of view, therefore, 
the argument is circular, a \emph{petitio principii}: what is to be proved has already been tacitly included among the premises.

\section{Symmetries and Conservation Laws}
One of the most important aspects of the concept 
of dynamical symmetry in modern physics is its 
close relation to so-called \emph{conservation laws}. 
These state that certain quantities remain unchanged 
throughout the given dynamical evolution. According 
to a celebrated theorem proved by the mathematician 
\textsc{Emmy Noether} in 1918, dynamical symmetries 
possessing certain further mathematical properties 
lead directly to conservation laws \cite{Noether:1918b}. 
For example, transformations of the form 
\eqref{eq:DarstellungE3} lead to conservation of 
momentum (associated with invariance under spatial 
translations $\vec x\mapsto\vec x+\vec a$) and 
angular momentum (associated with invariance under 
spatial rotations $\vec x\mapsto R\vec x$), whereas 
the velocity transformations \eqref{eq:BoostTransformation} 
lead to the conservation law governing the motion 
of the centre of mass. Finally, invariance under time 
translations \eqref{eq:Zeittranslation} gives 
conservation of energy. But not every dynamical 
symmetry automatically gives rise to a conservation 
law, as may be illustrated by the example of the 
scaling symmetry \eqref{eq:RaumZeitSkalierung}, 
since there is no further conservation law 
corresponding to it. The reason for this 
difference is an \emph{additional} mathematical 
property possessed by the symmetries 
\eqref{eq:DarstellungE3}, \eqref{eq:BoostTransformation}, 
and \eqref{eq:Zeittranslation}, but not by
\eqref{eq:RaumZeitSkalierung}, and which, according to \textsc{Noether}'s theorem, is necessary and sufficient 
for the existence of an associated conservation law. 
A precise account of this property would take us 
too far afield, but we can at least say the following: 
almost all dynamical laws of physics, in particular those of mechanics and all field theories of the fundamental 
interactions, can be formulated mathematically as a 
variational problem, according to which the dynamics 
is such that a certain quantity, called the 
\emph{action}, is stationary, meaning that its 
first-order variation vanishes. One says that the 
actual dynamical trajectory is a \emph{stationary point} 
of the action functional. The dynamical symmetries that lead to conservation laws are then those that additionally leave the action invariant, or at most change it by an additive constant, so that the differential of the action remains invariant. Not every symmetry needs to satisfy this condition, as one sees, for example, in the case of the scaling symmetry, which changes the action, and hence its differential, by a constant multiplicative factor. Nevertheless, these apparently special symmetries are the rule rather than the exception in physics, and have their own name: \emph{Noether symmetries}. 

\section{Spontaneous Symmetry Breaking}
\label{sec:SpontaneSymmetriebrechung}
\emph{Symmetry breaking} is usually understood as the 
dynamical transition from a state of higher symmetry to 
one of lower symmetry (or, in some circumstances, of no symmetry at all). In statistical physics and thermodynamics, such transitions between states of different symmetry play an important role because they are usually associated with so-called \emph{phase transitions}, as for example in the melting of a crystal, where the crystal symmetry is obviously completely destroyed. It is essential to remember that it is the symmetries of individual configurations or states that are broken, or more accurately \emph{reduced}. The symmetries of the underlying laws are unaffected. 

A similar distinction applies to spontaneous symmetry breaking, the phenomenon in which the 
\emph{ground state} of a dynamical theory---by 
which one usually means the state of lowest 
energy, provided such a state exists---does 
not possess the full dynamical symmetry of 
the theory, or at least is not invariant 
under all dynamical symmetry operations that 
preserve energy. A unique ground state must be invariant under energy-preserving dynamical symmetries: otherwise, a non-trivially acting group element could transform it into a different state of the same energy, contradicting uniqueness. In the case of spontaneous 
symmetry breaking, therefore, the ground 
state is not unique (one then calls the 
ground state \emph{degenerate}), and if the 
system is initially in an excited state, 
then, as energy is removed, it must 
\emph{spontaneously} choose one of the 
possible ground states, without this choice 
being constrained in any way by the 
dynamical laws. If the system is now in one 
of the ground states, this situation will in 
general be only metastable, since the system 
can pass into neighbouring ground states 
with an arbitrarily small expenditure of 
energy, provided these are continuously 
connected with the chosen state. Small 
fluctuations about this metastable 
equilibrium configuration will then satisfy 
a dynamical law that is derived from the full 
dynamics and whose symmetry group is precisely 
the subgroup of the full symmetry group that 
leaves the ground state invariant. 

This can be understood in terms of the 
distinction between symmetry and covariance 
described in 
Section\,\ref{sec:SymmetrieKovarianz}, for the 
transition from the full theory to the 
theory of small fluctuations about the 
ground state involves a decomposition of the dynamical quantities 
$X$ (of the full theory) into two subsets 
$X'$ and $H$, where $X'$ denotes the new 
dynamical quantities and $H$ denotes the 
new background quantities relative to which the 
fluctuations are measured.  All these quantities 
were dynamical in the full theory. What 
was formerly the dynamical symmetry group is 
now, in the first instance, only a covariance group. It is a symmetry group only insofar as its 
elements leave fixed the quantities $H$, 
which no longer participate in the dynamics. 
This is precisely the reverse of the process 
already discussed in Section\,\ref{sec:SymmetrieKovarianz}. 

This change of perspective, $X\leftrightarrow (X',H)$, has applications almost everywhere symmetry arguments are used. Let us again consider our 
example of the gravitational dynamics of $N$ 
point particles, now modified by embedding 
the system of point particles in an external 
stationary and homogeneous gravitational 
field of as yet unknown origin. This 
external field then possesses a preferred 
direction, so that the dynamical system of 
the $N$ point particles retains only a 
rotational symmetry whose axis of rotation 
is parallel to this preferred direction. But 
if this external field is the gravitational 
field of a very massive body whose dynamics 
remains essentially unaffected by that of 
the point masses, and if the latter are also 
at a very great distance from the centre of 
that body, so that its field may be regarded 
as homogeneous over the comparatively small 
extent of the $N$-particle system (one may 
imagine the $N$ particles, for example, to 
be in a laboratory and the external 
gravitational field to be that of the 
Earth), then by including the dynamical 
equations for this body one may recover the 
full dynamical symmetry group. In the 
original description, this would have been 
present only as a covariance group, provided 
one allowed its elements to act appropriately 
on the parameters of the ``external field''. 

Examples abound, extending all the way to cosmology, 
where, for example, the velocity field of 
cosmological matter and/or of the 3\,K 
background radiation singles out preferred 
reference frames that do not admit velocity 
transformations of the form 
\eqref{eq:BoostTransformation} (or their 
special-relativistic adaptation) as 
symmetries, provided these frames are not 
themselves part of the dynamical description 
but are understood only as background. If we tried to attain maximal symmetry by ultimately including the entire physical universe in the dynamical description, the price would be to render the symmetry claim wholly trivial through indistinguishability. As 
emphasised in Section\,\ref{sec:SymmUnterscheidbarkeit}, the physical content of a symmetry always involves changes that are observable in principle. One example is the claim that physics in terrestrial laboratories or in the Solar System is independent of their velocity, measurable in principle, relative to the reference frame in which the 3\,K cosmic background radiation appears isotropic.

\newpage

\end{document}